**Driving Education Advancements of Novice Drivers: A Systematic Literature Review**

**Anannya Ghosh Tusti**
*(Corresponding Author)*
Ingram School of Engineering
Texas State University
601 University Drive, San Marcos, Texas 78666
Email: gpk30@txstate.edu

**Anandi K Dutta**
Ingram School of Engineering
Texas State University
601 University Drive, San Marcos, Texas 78666
Email: myb43@txstate.edu

**Syed Aaqib Javed**
Ingram School of Engineering
Texas State University
601 University Drive, San Marcos, Texas 78666
Email: aaqib.ce@txstate.edu

**Subasish Das, Ph.D.**
Ingram School of Engineering
Texas State University
601 University Drive, San Marcos, Texas 78666
Email: subasish@txstate.edu

**ABSTRACT**

Most novice drivers are teenagers since many individuals begin their driving journey during adolescence. Novice driver crashes remain a leading cause of death among adolescents, underscoring the necessity for effective education and training programs to improve safety. This systematic review examines advancements in teen driver education from 2000 to 2024, emphasizing the effectiveness of various training programs, technology-based methods, and access barriers. Comprehensive searches were conducted across ScienceDirect, TRID, and journal databases, resulting in the identification of 29 eligible peer-reviewed studies. Thematic analysis indicated that technology-enhanced programs, such as RAPT, V-RAPT, and simulators, enhanced critical skills like hazard anticipation and attention management. Parental involvement programs, including Share the Keys and Checkpoints, demonstrated sustained behavioral improvements and adherence to Graduated Driver Licensing (GDL) restrictions. However, limited access due to socioeconomic disparities and insufficient long-term evaluations constrained broader effectiveness. The exclusion of non-U.S. studies and variability in research designs restricted the generalizability of findings. Integrated approaches that combine traditional education with innovative training tools and parental engagement appear promising for improving teen driver safety, with future research required to evaluate long-term effectiveness and ensure equitable access.

# INTRODUCTION
## Background on Teen Driver Safety

Most novice drivers are teenagers since many individuals begin their driving journey during adolescence, typically obtaining their learner's permits and provisional licenses during their mid-to-late teens. This period marks a crucial transition where young drivers must develop essential skills such as hazard perception, defensive driving, and situational awareness. Teen driving represents a critical area of road safety due to the heightened risks associated with novice drivers. Motor vehicle crashes remain a leading cause of death among teenagers in the United States, with young drivers aged 16 to 20 experiencing fatal crash rates nearly three times higher than their older counterparts (National Highway Traffic Safety Administration, 2022). In 2022 alone, over 2,000 teen drivers were involved in fatal crashes, underscoring the urgent need for effective educational and training programs (IIHS, 2023). Developmental immaturity, risk-taking behaviors, and inexperience contribute to these alarming statistics, further emphasizing the role of structured driver education in improving safety outcomes (AAA, 2018). Studies show that 16-year-olds, in particular, have the highest crash rates per mile driven, primarily due to limited driving exposure, distractions, and poor hazard perception (Hossain et al., 2023, 2022; Lee et al., 2011; McCartt et al., 2009). Earlier approaches to driver education focused primarily on teaching basic vehicle operation and road rules, often through classroom instruction and limited supervised practice. However, advances in educational methodologies, such as Graduated Driver Licensing (GDL) systems and simulation-based training, have significantly enhanced the preparedness of teenage drivers. GDL systems, for instance, have been widely implemented to reduce exposure to high-risk situations, with phases that include supervised learning permits, restricted licenses, and eventually full driving privileges (Foss et al., 2012; Williams, 2017). Despite these advancements, disparities in access to driver education persist, particularly for underserved communities, rural populations, and economically disadvantaged families, resulting gaps in equitable road safety (Curry et al., 2012; Ryerson et al., 2022, 2024).

## The Importance of Driver Education and Training

Comprehensive driver education and training programs are pivotal in reducing the incidence of crashes among teenage drivers. Evidence suggests that structured programs focusing on hazard anticipation, risk assessment, and safe driving behaviors can significantly lower crash risks. For instance, in a study Curry et al. (2015) demonstrated that teens who completed rigorous driver education programs exhibited safer driving patterns and fewer traffic violations compared to those without formal training. Additionally, advanced training tools, such as virtual reality simulations and PC-based modules like DriverZED, have been shown to improve situational awareness and decision-making skills in novice drivers (Goode et al., 2013; Thomas et al., 2012). An analysis by Mayhew et al. (2024) found that driver education programs incorporating practical and cognitive training components achieved greater reductions in risky driving behaviors compared to traditional classroom-only models. These findings align with the work of Curry et al., (2015) and Gesser-Edelsburg and Guttman, (2013), which emphasized the role of parental involvement in reinforcing safe driving habits. However, the effectiveness of driver education programs often varies due to inconsistencies in curriculum standards, accessibility, and implementation. States with mandatory driver education requirements report better outcomes, yet only 23 states currently enforce these mandates for all drivers under 18 (Thomas et al., 2012). The mandatory driving education requirements differ based on state regulations. Moreover, socioeconomic barriers, such as the high costs of private driving schools, exacerbate inequities in access to training. Addressing these challenges requires a multifaceted approach that integrates education, policy reform, and community engagement to ensure all teens have the opportunity to become safe and competent drivers.

**Challenges in Equitable Access to Driver Education**
Despite its importance, access to driver education remains inequitable across the United States. Underserved communities, including low-income families, immigrants, and rural populations, face significant barriers to participating in formal training programs. Income and geographic disparities also play a role, with rural areas often lacking nearby training facilities, further limiting accessibility (FHWA, 2022). A comprehensive study by (Masten and Foss, 2010) revealed that driver education programs in rural areas often have limited access to high-quality instructors and modern training tools, leading to less effective outcomes. Furthermore, variations in state policies create inconsistencies in driver education requirements. For instance, as of 2023, Vermont lacks nighttime driving restrictions for novice drivers[1], and New Hampshire does not enforce a minimum holding period for learner's permits (GAO, 2010). Such variations lead to disparities in preparedness and safety outcomes among teenage drivers nationwide.

**Rationale for the Review**
Given the critical role of driver education in shaping safe driving behaviors, this systematic review aims to synthesize existing research on teen driving education and training programs. By identifying effective strategies, evaluating their impact on crash rates, and highlighting barriers to implementation, this review seeks to inform policy decisions and promote equitable access to driver education. Furthermore, it addresses gaps in literature by focusing on integrating innovative technologies, such as simulation-based training and gamified learning, into traditional curriculum.

**Research Objectives and Questions**
The primary objective of this review is to systematically evaluate advancements in teen driver education programs, emphasizing their effectiveness, the role of innovative training methods, and the barriers to equitable access. This review also aims to identify evidence-based strategies for reducing crash risks and promoting safe driving behaviors among teen drivers. Specific research questions include:
1. How effective are both traditional and emerging driver education programs in reducing crash risks and enhancing critical driving skills such as hazard perception, risk awareness, and attention management among novice teen drivers?
2. How do innovative, technology-enhanced training methods, including simulator-based programs, risk awareness perception training (RAPT), and virtual reality, improve safety outcomes and cognitive skill retention for teen drivers?
3. What role do socioeconomic, geographic, and policy-related barriers play in limiting access to quality driver education, and how can these barriers be mitigated to promote equitable access across diverse communities?

---

[1] Teen drivers are often classified as novice drivers due to their limited experience behind the wheel. However, the term 'novice driver' is not exclusive to teenagers. Individuals beyond their teenage years who have little to no prior driving experience also fall into this category. Whether a driver is in their twenties, thirties, or even older, the challenges associated with being a novice remain similar, such as developing hazard perception skills, mastering vehicle control, and gaining confidence in diverse traffic conditions.

4. How can sustained parental engagement and structured supervision improve compliance with GDL restrictions and implement long-term safe driving practices among teens?

By systematically addressing these questions, this review aims to provide actionable insights for policymakers, educators, and community organizations working to improve teen driver safety. Addressing these challenges can significantly contribute to reducing teen crash rates and fostering a culture of safety on the roads.

**METHODOLOGY**

This study followed the Preferred Reporting Items for Systematic Reviews and Meta-Analysis (PRISMA) guidelines to ensure a structured, transparent, and reproducible approach to reviewing the existing literature (Moher et al., 2009). The PRISMA checklist was followed step by step to guide the selection, screening, and analysis of relevant studies. The objective of this review was to systematically identify and synthesize research focused on advancements in education and training systems for novice teen drivers. To achieve this, a comprehensive literature search was conducted across multiple academic databases and reputable journal websites, ensuring the inclusion of high-quality, peer-reviewed studies.

A systematic search protocol was implemented to locate relevant studies published between January 2000 and December 2024, with the final search conducted in January 2025. The selected databases and sources included ScienceDirect, Transportation Research International Documentation (TRID), and journal publisher websites that host transportation safety and education-related research. The keywords used in the search process included "novice" OR "teen" AND "education" OR "training", ensuring the retrieval of studies relevant to driver education. The search was applied to the title, abstract, and keywords of articles, and no language restrictions were applied. However, later studies were not available in English. Initially, studies were collected from 1972 to 2024, from which studies spanning 2000-2024 were filtered according to the protocol followed in this study.

The selection of studies was guided by well-defined inclusion and exclusion criteria. Studies were included if they were published within the specified time range and focused on teen driver education, training programs, learning methodologies, or technology-based driving education. Research that assessed the effectiveness of different educational interventions or provided empirical data, systematic reviews, or meta-analyses was prioritized. In contrast, studies published before 2000, those that did not focus on teen driver education or training, opinion pieces, non-peer-reviewed sources, duplicate publications, and studies lacking sufficient methodological details were excluded. Following the search process, all identified studies were documented in an MS Excel spreadsheet, capturing key details such as document type (journal article, report, conference paper), publication year, author names, title, abstract, publisher information, DOI, and URL. This structured approach ensured a clear organization of relevant literature before the screening phase.

The study selection process occurred in two phases. Initially, two independent reviewers screened the titles and abstracts of all retrieved studies to identify those that met the inclusion criteria. Studies that clearly did not align with the scope of this review were excluded at this point. In the second phase, the remaining studies underwent a full-text review to confirm their relevance and methodological quality. Any discrepancies between the reviewers were resolved through discussion or, when necessary, by consulting a third reviewer. The entire selection process is illustrated in a PRISMA Flow Diagram, showing the number of records identified, duplicates removed, studies screened, full-text articles assessed for eligibility, and the final set of studies included in the review.

Once the studies were selected, data synthesis and analysis were conducted using both qualitative and quantitative approaches. Thematic synthesis was employed to identify common themes, advancements, and challenges in teen driver education. Additionally, where applicable, quantitative findings were summarized, particularly those measuring the effectiveness of different educational interventions using statistical indicators. The synthesis of results helped draw meaningful conclusions regarding the most effective educational strategies for improving teen driving behavior and safety outcomes.

Despite efforts to ensure a rigorous and unbiased review, certain limitations were acknowledged. Publication bias may have affected the findings, as studies reporting negative or non-significant results may be underrepresented in published literature. Selection bias was another concern, as the review was limited to specific academic databases and reputable journals. Additionally, language bias was a potential limitation, as studies not available in English were excluded. To mitigate these biases, two independent reviewers conducted the screening process, and any disagreements were resolved through consensus or consultation with an expert. Since this study is a systematic review of existing literature, no ethical approval was required. However, to maintain academic integrity, all sources were properly cited and referenced according to standardized citation formats. Only the studies selected by PRISMA protocol were described in the RESULTS section where the INTRODUCTION section contained other studies for presenting a better understanding of the need for this study.

**RESULTS**
**Study Selection**
The systematic review on advancements in teen driver education and training targeted studies published between 2000 and 2024. An extensive search was conducted across multiple databases, including ScienceDirect, TRID, and reputable journal websites, as well as through citation tracking. This effort identified 238 records through databases and an additional 6 through citation searching. Following the removal of 14 duplicates, 2 ineligible records flagged by automation tools, and 42 records published prior to 2000, 180 studies were eligible for title and abstract screening. This process excluded 4 studies due to language limitations, leaving 176 reports for full-text retrieval. However, 54 reports could not be retrieved, leaving 122 studies for detailed eligibility assessment. During this stage, 93 studies were excluded for the following reasons: 27 focused on novice adult drivers, 51 on non-U.S. contexts, and 18 did not directly address teen driver education. Additionally, 6 studies identified through citation searches were screened, with 2 excluded for focusing on novice adult drivers. Ultimately, 29 studies met the inclusion criteria and were synthesized in the review. The rigorous selection process ensured a robust and high-quality dataset, reflecting the diversity of approaches and challenges in teen driver education programs. Figure 1 represents the study selection using PRISMA method.

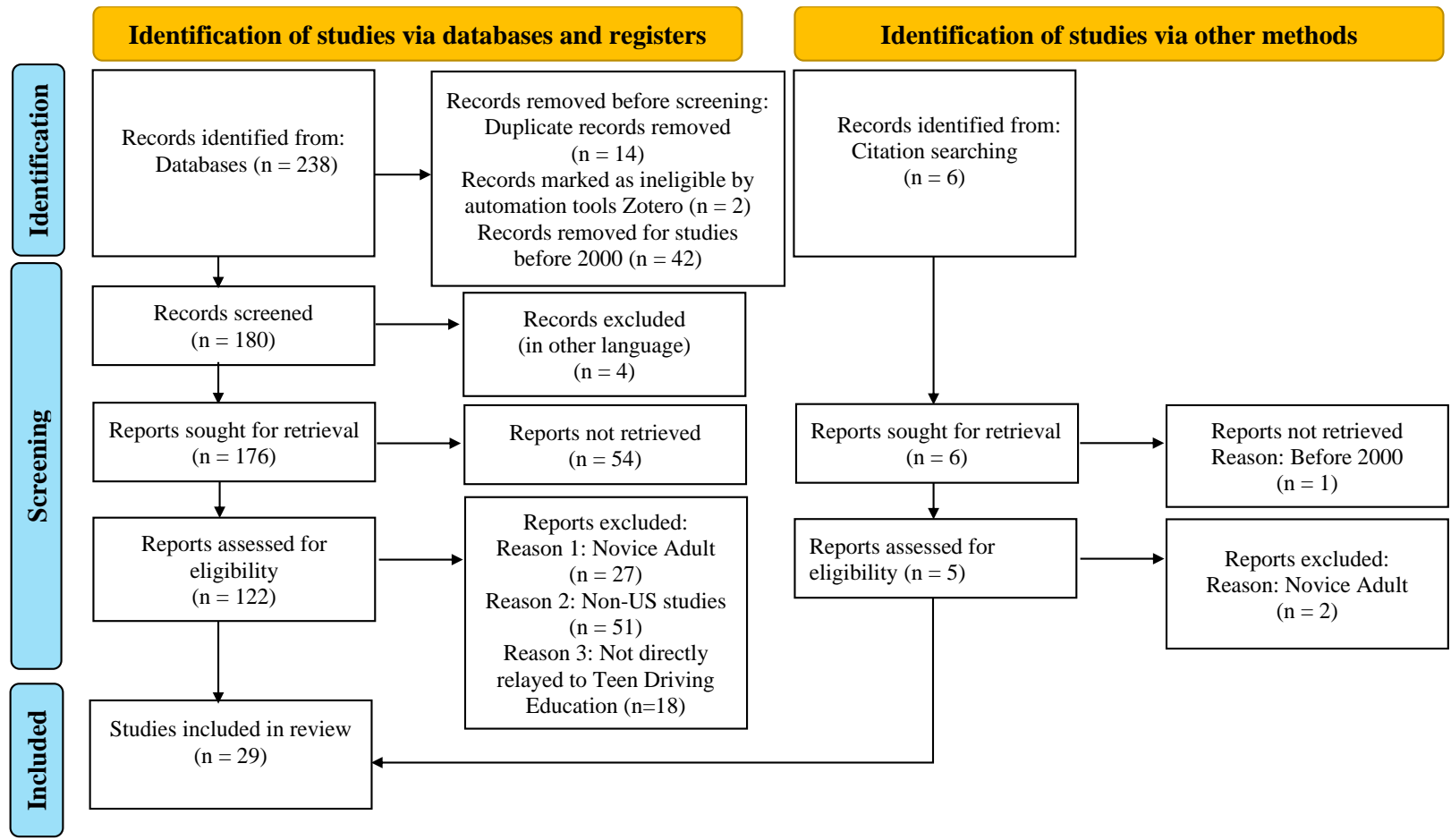

**Figure 1 PRISMA flowchart for study selection**

**Study Characteristics**
The systematic review included a total of 29 studies categorized into relevant themes and training programs, as shown in Figures 1 and 2. These studies were conducted between 2000 and 2024 and aimed to improve teen driver safety by addressing various aspects of driver education and training. Studies examined traditional driver education programs, technology-enhanced methods, risk awareness, hazard perception training, and the influence of psychological, cognitive, and social factors on driver behavior. Some studies were present in more than one category due to the relevance of the information of that particular category. The studies employed diverse methodologies, including experimental studies with driving simulations and eye-tracking (McDonald et al., 2015b; Pradhan et al., 2005), policy reviews (Alderman et al., 2018; NHTSA, 2009), and longitudinal evaluations (Glassman et al., 2024; Thomas et al., 2016). The objectives ranged from evaluating the impact of specific training programs like RAPT, ACCEL, and EDTS on hazard anticipation and perception to assessing broader educational initiatives such as high school driver education and GDL programs. The target populations primarily consisted of novice teen drivers aged 15 to 21 years, with a few studies focusing on parental involvement and peer influence. Training programs were implemented across various formats, including PC/tablet-based training (e.g., Zafian et al., 2016), simulator-based programs (e.g., Pollatsek et al., 2011), and on-road evaluations (e.g., Knezek et al., 2018). The studies offered valuable insights into the effectiveness of these interventions, highlighting the importance of tailored hazard perception training, active parental involvement, and continual skill reinforcement to reduce crash rates and enhance overall safety.

The data were systematically compiled in a table summarizing key study characteristics such as year of publication, methodology, training program type, and age range of participants. In cases where information was missing, efforts were made to contact the original authors to ensure accuracy and comprehensiveness in reporting. This approach enables readers to evaluate the relevance and validity of the findings in the context of teen driver education and safety programs. Table 1 presents a summary of the reviewed papers. Figure 2 depicts the classification of studies based on the type of training programs and interventions examined in the selected research. Among the 29 studies reviewed, several falls into multiple categories depending on the scope and focus of their research. The classification identifies five major categories of driver education and training: Traditional Driver Education Programs, Technology-Enhanced Training Methods, Risk Awareness and Hazard Perception Training, Psychological and Cognitive Factors in Teen Driver Education, and Parental and Peer Influence on Teen Driving Education. The Technology-Enhanced Training Methods category reflects the growing reliance on digital tools, simulations, and online learning to improve driver training outcomes. The Risk Awareness and Hazard Perception Training category emphasizes research aimed at enhancing situational awareness and decision-making skills to reduce crash risks. Studies under Psychological and Cognitive Factors in Teen Driver Education explore the effects of cognitive abilities, attention, and learning behaviors on driver training effectiveness. Finally, Parental and Peer Influence on Teen Driving Education highlights the role of social factors in shaping young drivers' behavior and risk perception.

The analysis of driver training programs over different time periods is illustrated in Figure 3. This stacked bar chart categorizes training methods into five distinct types: GDL, Driver Education and Licensing Standards, Hazard Training, Simulator-based Training, PC/Tablet-based Training, and School Training Programs. The graph highlights the evolution and distribution of these training methods across different time frames. A noticeable peak in research interest is observed in 2016–2020, with a significant rise in studies focusing on PC/Tablet-based Training and School Training Programs, indicating a shift toward technology-driven and structured learning approaches. A slight decline is seen in 2021–2024, though Simulator-based Training and PC/Tablet-based Training remain prevalent. This trend suggests that while traditional training programs continue to play a role, technology-driven methods such as simulation-based learning and digital platforms are becoming dominant. The Keyword Co-occurrence Network (KCN) in Figure 4, generated using VOSviewer, visualizes the dominant themes in the selected papers by extracting keywords from titles, abstracts, and citation contexts and mapping their co-occurrence relationships. The distance between keywords is inversely proportional to their relatedness, while clusters represent groups of frequently co-occurring terms, with the area covered indicating the strength of their association. The visualization reveals that 'driver' is the most dominant key term, followed by 'training', 'teen driver', 'training program', and 'crash'. Keywords like 'novice driver', 'risk awareness', and 'simulator training' are tightly connected, indicating strong interrelationships in research, whereas terms like 'situation awareness' and 'policy statement' are farther apart, reflecting lower co-occurrence rates. This network highlights the key research trends in driver training, risk awareness, and crash prevention, providing insights into emerging focus areas in the field.

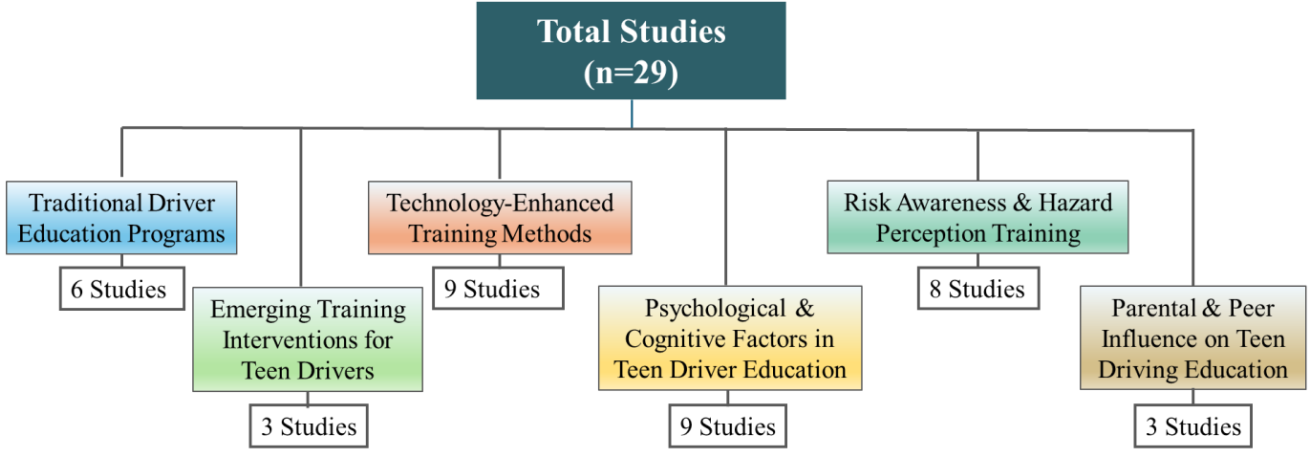

**Figure 2 Systematic classification of studies into relevant categories**

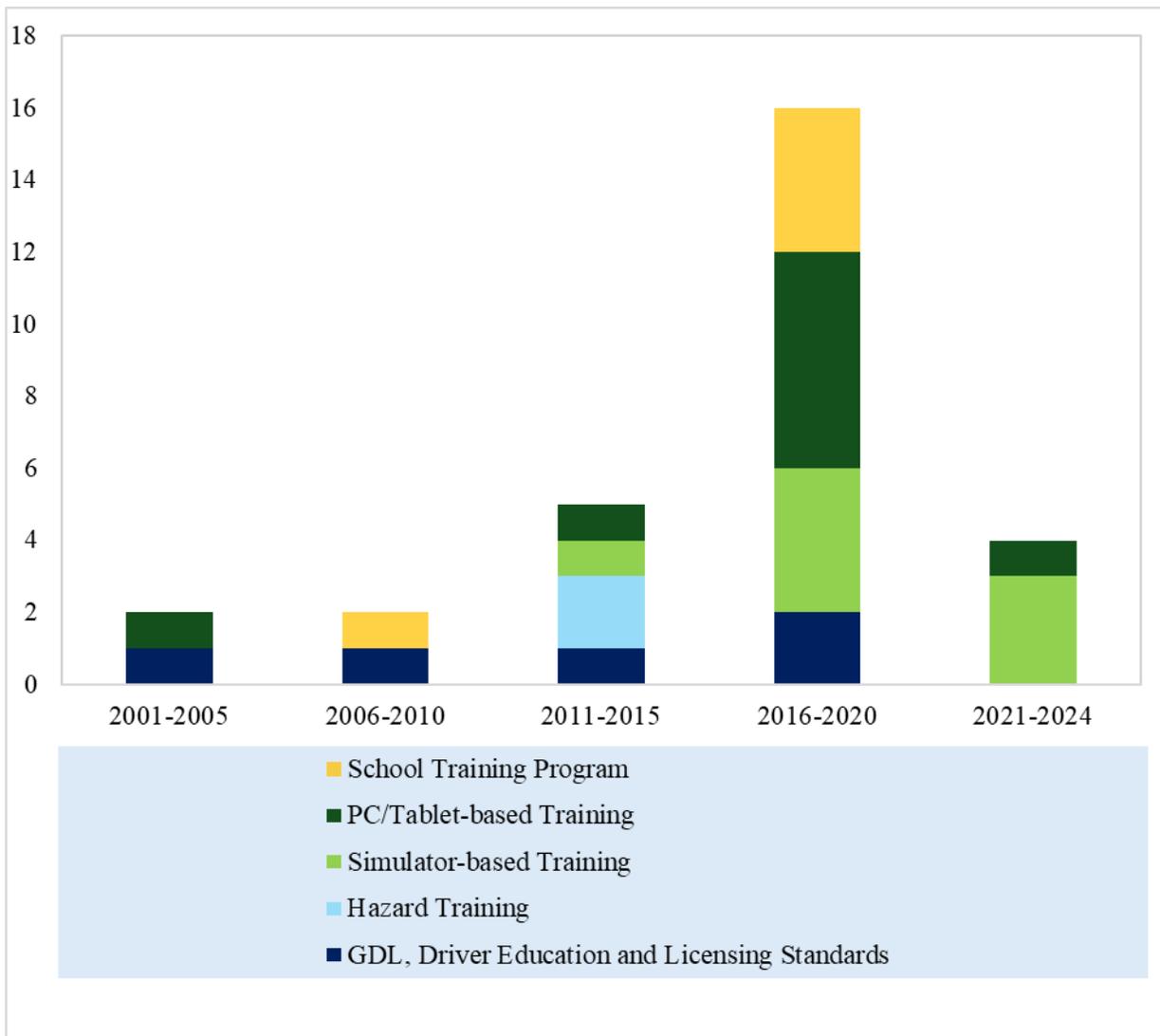

**Figure 3 Systematic classification of studies according to training/program studied**

**Table 1 A summary of the reviewed studies**

| Study Name | Objective of the Study | Methodology Followed | Education/Training Program Studied | Year | Age Range |
|---|---|---|---|---|---|
| (NHTSA, 2009) | Establish standardized guidelines for novice teen driver education. | Policy reviews and expert consensus approach. | Novice driver education standards | 2009 | Teen drivers (16-19) |
| (Pradhan et al., 2005) | Assess the impact of PC-based risk awareness training on novice drivers. | Experimental study with driving simulation and eye-tracking. | PC-based risk awareness training | 2005 | 16-19 years |
| (Edwards, 2001) | Examine the role of driver education in the licensing process. | Review of existing standards, policy guidelines, and expert consultations. | Novice driver education and licensing standards | 2001 | 16-18 years |
| (Fisher and Dorn, 2016) | Identify best practices in training novice teen drivers. | Literature review and policy recommendations. | Novice driver education programs | 2016 | 16-19 years |
| (Lonero and Mayhew, 2010) | Review the effectiveness of large-scale driver education programs. | Meta-analysis of past evaluations and program assessments. | High school and commercial driver education programs | 2010 | 16-18 years |
| (Pollatsek et al., 2011) | Investigate the role of driving simulators in training and evaluation. | Review of simulator-based training effectiveness. | Simulator-based driver training | 2011 | 16-21 years |
| (Allen et al., 2012) | Assess long-term effectiveness of simulator training on novice drivers. | Longitudinal study with simulator-based training and post-test evaluation. | Simulator-based hazard training | 2012 | Novice drivers (16-18 years) |
| (McDonald et al., 2015a) | Systematic review of hazard anticipation programs for young drivers. | Literature review covering multiple hazard anticipation training programs. | Multiple hazard anticipation training programs | 2015 | Young drivers (<21 years) |
| (Shell et al., 2015) | Compare crash and violation rates of driver-educated teens vs. non-educated teens. | Descriptive epidemiological study of a teen driver cohort. | Traditional driver education | 2015 | 16-18 years |
| (McDonald et al., 2015b) | Assess the impact of RAPT-3 training on hazard perception at left-turn intersections. | Randomized controlled trial using a driving simulator. | RAPT-3 risk awareness training | 2015 | 16-18 years |
| (Campbell et al., 2016) | Examine whether driving simulator training reduces crashes and violations. | Randomized study with pre- and post-training assessments. | Simulator-based driver training | 2016 | Teen drivers |

| Study Name | Objective of the Study | Methodology Followed | Education/Training Program Studied | Year | Age Range |
|---|---|---|---|---|---|
| | | | | | (16-18 years) |
| (Thomas et al., 2016) | Assess long-term impact of RAPT on crash rates and violations. | Large-scale study tracking 5,251 young drivers over 12 months. | RAPT-based risk perception training | 2016 | 16-18 years |
| (Zafian et al., 2016) | Evaluate the real-world impact of a tablet-based driver training intervention. | Field study comparing tablet-trained and placebo-trained drivers. | Tablet-based driver training | 2016 | Novice teen drivers |
| (Fisher et al., 2017) | Evaluate the impact of the ACCEL training program on novice drivers' hazard anticipation, mitigation, and attention maintenance skills | Driving simulator evaluation, two-part experimental study | ACCEL Training Program | 2017 | 16-19 |
| (Muttart et al., 2017) | Assess effectiveness of ACT training on novice drivers' hazard mitigation behavior | Driving scenarios with experimental and control groups | ACT Training | 2017 | 16-17 |
| (Agrawal et al., 2017) | Evaluate effectiveness of V-RAPT training using virtual reality | Driving simulator evaluation, eye movement tracking | V-RAPT | 2017 | 16-21 |
| (Thomas et al., 2017) | Evaluate an updated version of RAPT in improving hazard perception skills in novice drivers | On-road evaluation with pre- and post-training assessments | RAPT | 2017 | 16-19 |
| (Knezek et al., 2018) | Evaluate the STK program's impact on parental involvement and teen driver safety | Pre-, post-, and follow-up surveys across multiple phases | Share the Keys (STK) | 2018 | 16-18 |
| (Ahmadi et al., 2018) | Investigate the short- and long-term impact of EDTS on hazard perception | Experimental study with on-road evaluation, eye movement tracking | Engaged Driver Training System (EDTS) | 2018 | 16-19 |
| (Unverricht et al., 2019) | Examine whether the FOCAL training improves both attention maintenance and multitasking in young drivers | Driving simulator with in-vehicle secondary tasks, attention tracking | FOCAL Program | 2019 | 18-24 |
| (Reyes and O'Neal, 2020) | Assess effects of two hazard anticipation training programs using driving simulators | Simulated hazard scenarios with eye-tracking and performance measurement | PALM and ACCEL Programs | 2020 | 15-16 |
| (Wang et al., 2020) | Evaluate the crash reduction impact of advanced driver training programs | Randomized controlled trial with survival analysis over 550 days | Advanced Driver Training Program | 2020 | 16-19 |

| Study Name | Objective of the Study | Methodology Followed | Education/Training Program Studied | Year | Age Range |
| --- | --- | --- | --- | --- | --- |
| (Yahoodik and Yamani, 2020) | Examine how RAPT training affects attentional control in dynamic driving scenarios | Driving simulator with latent hazard scenarios and dynamic stimuli | RAPT Program | 2020 | 16-19 |
| (Plumert et al., 2021) | Evaluate long-term hazard anticipation performance in young drivers | Driving simulator with longitudinal assessments | PALM and ACCEL Programs | 2021 | 16-19 |
| (Glassman et al., 2024) | Examine the impact of a second dose of ACCEL training on hazard anticipation performance | Longitudinal simulator study with eye tracking | ACCEL Program | 2024 | 16-18 |
| (Unverricht et al., 2018) | Conduct a meta-analysis on training studies aimed at improving young drivers' latent hazard anticipation skills | Meta-analysis of 19 peer-reviewed studies using driving simulators and on-road evaluations | Multiple programs (e.g., RAPT, SAFE-T) | 2018 | 16-25 |
| (O'Neill, 2020) | Review the effectiveness of formal driver education programs in reducing crashes | Historical review of randomized control trials and longitudinal studies | High School and Formal Driver Education Programs | 2020 | 16-19 |
| (Alderman et al., 2018) | Highlight the unique risks faced by teen drivers and outline strategies for safety | Policy review, observational studies, and data analysis | GDL and Parental Programs | 2018 | 15-19 |
| (Das et al., 2019) | Assess teen drivers' understanding of factors contributing to car crashes | Survey data analysis using Taxicab Correspondence Analysis (TCA) | Teens in the Driver Seat (TDS) | 2019 | 13 to 19 years old (high school students) |

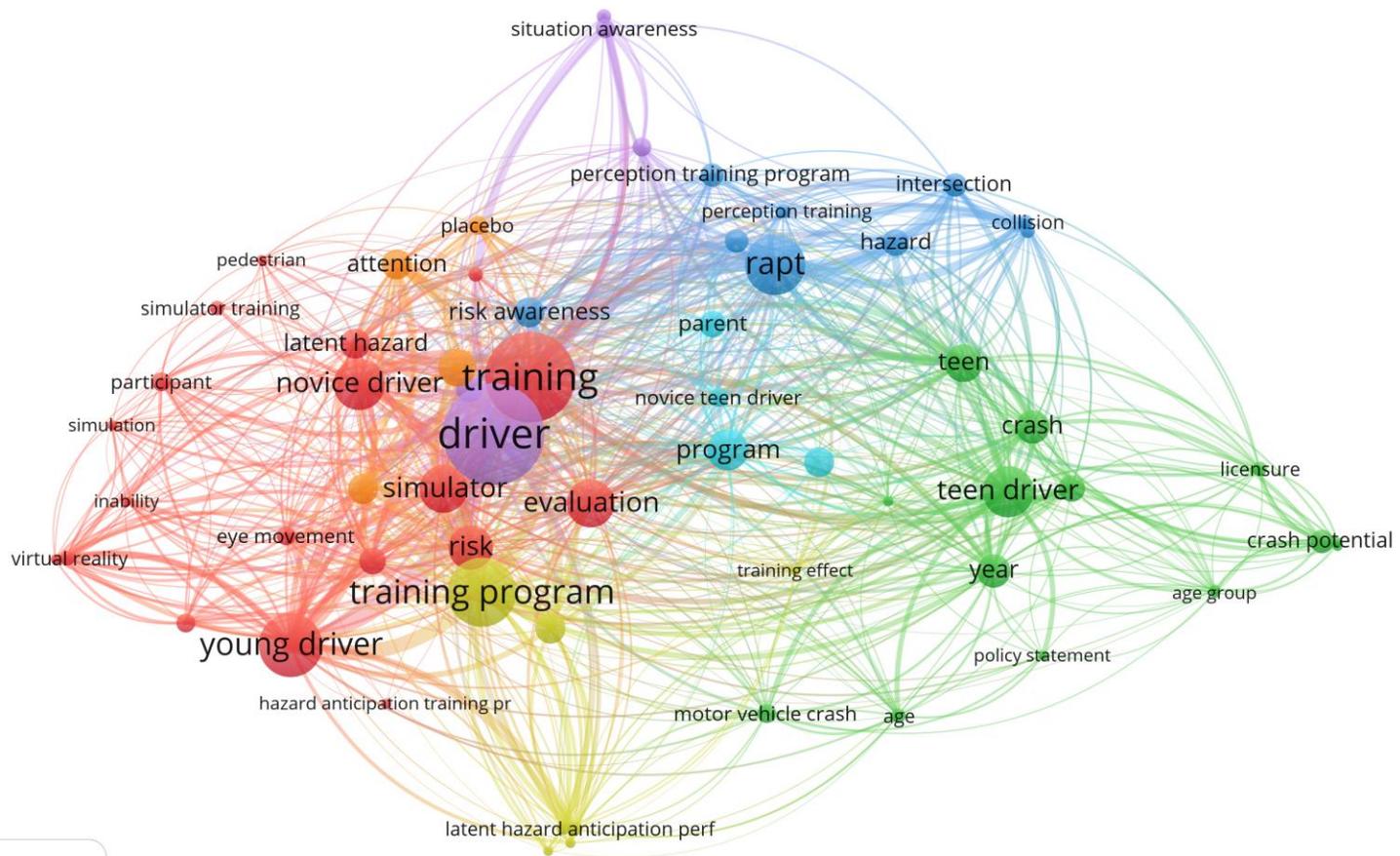

**Figure 4 Keyword co-occurrence network**

**Synthesis of the Results**
**Definition of Novice Driver**
A novice driver is typically defined as a newly licensed individual with limited independent driving experience, often within the first few months to two years of driving. This category primarily includes teenagers who have recently acquired a provisional or intermediate license. For teen drivers, novice status is typically defined and categorized within the framework of GDL systems, which aim to gradually increase driving privileges as teens gain experience and maturity. These categories include:

*Learner Teen Drivers*
Teens who hold a learner's permit and must drive under the supervision of a licensed adult, usually a parent or instructor. During this phase, they focus on acquiring basic vehicle control and road safety knowledge through practice and formal driver education programs (Knezek et al., 2018; Reyes and O'Neal, 2020).

*Intermediate or Provisional Teen Drivers*
Teens who have earned a restricted license and are allowed to drive independently but under certain limitations, such as restrictions on nighttime driving and the number of teen passengers. This stage is critical, as studies show that crash risks remain high due to limited hazard anticipation and risk management skills (Knezek et al., 2018).

*Newly Fully Licensed Teen Drivers*
These are teens who have completed the GDL process and have unrestricted driving privileges but still lack extensive experience. Research indicates that novice teen drivers in this stage are at heightened risk for crashes, particularly in complex or high-pressure driving situations (Reyes and O'Neal, 2020). Research highlights that novice drivers are at a significantly higher risk of crashes due to underdeveloped skills in hazard anticipation, risk awareness, and attention management.

**Effectiveness of Traditional Driver Education Programs**
Studies have evaluated various traditional driver education programs designed to train novice teen drivers, offering insights into their effectiveness in reducing crashes and improving driver behavior. These programs typically provide a combination of classroom instruction and behind-the-wheel training, often embedded within high school curricula or delivered by private driving schools. One of the earliest systematic approaches to teen driver education was introduced through high school driver education programs in the United States in the 1930s. These programs became highly prevalent by the 1960s, offering a standardized curriculum consisting of approximately 30 hours of classroom instruction and six hours of behind-the-wheel training. The DeKalb County Driver Education Project in the 1980s was a landmark study, testing a comprehensive program involving in-depth classroom instruction and hands-on driving sessions. The study found initial reductions in crash rates but also highlighted concerns regarding early licensure leading to increased exposure to crashes (Edwards, 2001; Lonero and Mayhew, 2010; O'Neill, 2020).

A major initiative by the National Highway Traffic Safety Administration (NHTSA) introduced the Novice Teen Driver Education and Training Administrative Standards, a framework developed to guide states in establishing high-quality, consistent driver education programs. The standards emphasized a phased approach to learning, integration with GDL systems, and alignment with evidence-based practices to improve crash prevention(Lonero and Mayhew, 2010; NHTSA, 2009).

Key elements included the recommendation of a 45-hour classroom curriculum and 10 hours of on-road instruction, often modeled after curricula developed by organizations like the American Driver and Traffic Safety Education Association (ADTSEA) (NHTSA, 2009). The AAA Foundation for Traffic Safety also conducted large-scale evaluations of driver education programs, emphasizing the need for more stringent evaluations to gauge long-term effectiveness. Despite high public expectations for these programs to improve safety outcomes, evaluations revealed mixed results, with only moderate improvements in knowledge and short-term crash reductions (Lonero and Mayhew, 2010). In some cases, programs demonstrated improvements in skills but struggled to influence risky behaviors such as speeding and distracted driving (O'Neill, 2020).

A recent study by Das et al. (2019) provides further insight into the relationship between teen drivers' understanding of crash risks and their driving experience across various licensing stages. Using data from approximately 88,000 Texas high school students surveyed under the Teens in the Driver Seat (TDS) program, the study revealed that teen perceptions of crash potentials vary based on their gender and stage of licensure. The research employed Taxicab Correspondence Analysis (TCA) to highlight significant differences in teens' awareness of driving risks, underscoring the value of targeted interventions at each licensing stage. Findings from this study reinforce the need for a phased and tailored approach to driver education to address developmental differences among teen drivers. In the state of Nebraska, an alternative training approach under the GDL system required teens to either complete a certified driver education course or accumulate 50 hours of supervised driving with a licensed adult. Research comparing these two methods found that teens who underwent formal driver education had significantly fewer crashes and traffic violations in their first two years of driving, supporting the program's utility within the GDL framework (Shell et al., 2015). Internationally, programs in countries like Sweden, Finland, and Luxembourg implemented second-stage driver education, which required newly licensed drivers to complete additional training after a probationary period. Evaluations of these programs noted some positive safety outcomes, particularly in reinforcing hazard perception and safer driving practices (Lonero and Mayhew, 2010).

Despite advancements in program design, criticisms persist. Studies by O'Neill (2020) argue that the historical emphasis on classroom-based training and early licensure may have inadvertently contributed to higher crash rates among teen drivers by accelerating their independence without adequately addressing maturity-related risk factors  The review also highlighted a lack of significant improvements in crash outcomes across various educational contexts, suggesting the need for alternative approaches focusing on behavioral change rather than purely technical skills (Lonero and Mayhew, 2010; O'Neill, 2020). These findings collectively emphasize that while traditional driver education programs provide essential foundational skills, their effectiveness in reducing long-term crash risks remains limited. Enhancements such as incorporating risk awareness, stronger parental involvement, and integration with broader licensing policies are necessary to improve the safety outcomes of novice teen drivers. The summary of the key findings for studies related to traditional driver education programs is presented in Table 2.

**Table 2. Summary of the Effectiveness of Traditional Driver Education Programs.**

| Study | Key Findings/Summary |
|---|---|
| O'Neill (2020) | Highlighted that traditional high school driver education often led to earlier licensure, increasing exposure to crash risks rather than reducing them. Questioned its effectiveness without complementary measures. |

| Shell et al., (2015) | Criticized inconsistencies in traditional driver education programs across states. Advocated for integrating these programs with graduated licensing systems to improve outcomes. |
|---|---|
| (Lonero and Mayhew, 2010) | Found a significant decline in participation in traditional high school programs due to funding cuts and lack of evidence for safety benefits. Recommended parental involvement and supervised practice as complementary measures. |
| (NHTSA, 2009) | Identified limitations in traditional driver education, such as a lack of focus on behavioral change. Stressed the need for enhanced evaluation methods to measure long-term safety impacts. |
| (Edwards, 2001) | Demonstrated that teens taking driver education within a structured GDL system had lower crash and violation rates compared to those relying solely on supervised practice, suggesting that integration with modern systems enhances effectiveness. |
| (Das et al., 2019) | Found that teen drivers' understanding of crash risks varies by gender and licensing stage, reinforcing the need for phased, targeted driver education programs. |

**Emerging Training Interventions for Teen Drivers**

In response to the limited success of traditional driver education programs, various emerging training interventions have been developed to enhance hazard perception, risk management, and overall driving behavior among teen drivers. These interventions utilize simulations, rule-based methods, and experiential learning to address specific challenges faced by novice drivers. One notable intervention is the Anticipate, Control, and Terminate (ACT) training program, which builds upon the previously developed RAPT. The ACT program emphasizes key skills such as hazard anticipation, speed management, and lane positioning through scenarios involving curves, intersections, and straight road segments. Studies have shown that ACT-trained drivers demonstrate improved anticipatory glances and earlier speed adjustments when approaching curves, resulting in better crash avoidance outcomes (Muttart et al., 2017).

The advanced driver training programs (ADTPs) provide hands-on practice for vehicle handling in controlled environments. These programs often include defensive driving exercises such as skid control, emergency stopping, and evasive maneuvers. Despite their popularity and perceived effectiveness among participants and parents, evaluations have yielded mixed results. A study involving a large sample of traffic offenders in North Carolina found that ADTP participants showed no significant crash reduction benefits compared to those who attended traditional classroom programs (Wang et al., 2020). Critics argue that such programs may inadvertently increase driver overconfidence without sufficiently improving core driving skills. Another emerging intervention focuses on hazard mitigation at specific high-risk locations, such as curves. A simulator-based study evaluated the effectiveness of training programs designed to improve speed and glance behaviors on curves. The study revealed that drivers who underwent targeted training made more appropriate speed adjustments and glances at potential hazards, significantly enhancing their hazard mitigation capabilities compared to placebo-trained drivers (Muttart et al., 2017). Internationally, second-stage driver training programs in countries like Finland and Sweden aim to reinforce safe driving practices post-licensure. These programs require newly licensed drivers to participate in additional training sessions that emphasize hazard perception, risk awareness, and crash scenario simulations (Fisher and Dorn, 2016).

Despite these advancements, the overall effectiveness of emerging interventions remains a subject of ongoing research. There is a recognized need for large-scale, controlled evaluations to determine the long-term impact of these programs on crash rates and risky driving behaviors. The Transportation Research Board has emphasized the importance of further studies to assess the potential benefits and limitations of these new training models (Wang et al., 2020). Table 3 represents the key findings of the studies related to emerging training interventions for teen drivers.

**Table 3. Emerging Training Interventions for Teen Drivers.**

| Study | Key Findings/Summary |
|---|---|
| Muttart et al. (2017) | Evaluated the ACT training program for hazard mitigation on curves. ACT-trained drivers exhibited better speed control and anticipatory glances, reducing crash likelihood in simulated driving conditions. |
| Wang et al. (2021) | Assessed an Advanced Driver Training Program (ADTP) aimed at young traffic offenders. Found no significant difference in crash reduction compared to traditional classroom-based programs, raising concerns about the effectiveness of hands-on short-term interventions. |
| (Fisher and Dorn, 2016) | Improving hazard anticipation and cognitive awareness, with programs like RAPT and PALM showing significant short-term improvements in risk detection and reaction skills. |

**Technology-Enhanced Training Methods**
A range of innovative technology-driven training programs has been developed to address the critical skills gap that often leads to higher crash rates among novice teen drivers. These programs use simulations, virtual reality, and tablet-based interactive modules to teach risk awareness, hazard anticipation, and decision-making. Table 4 and Figure 5 present information about the technology-enhanced training methods.Figure 5 Simulation and PC/Tablet-based programs

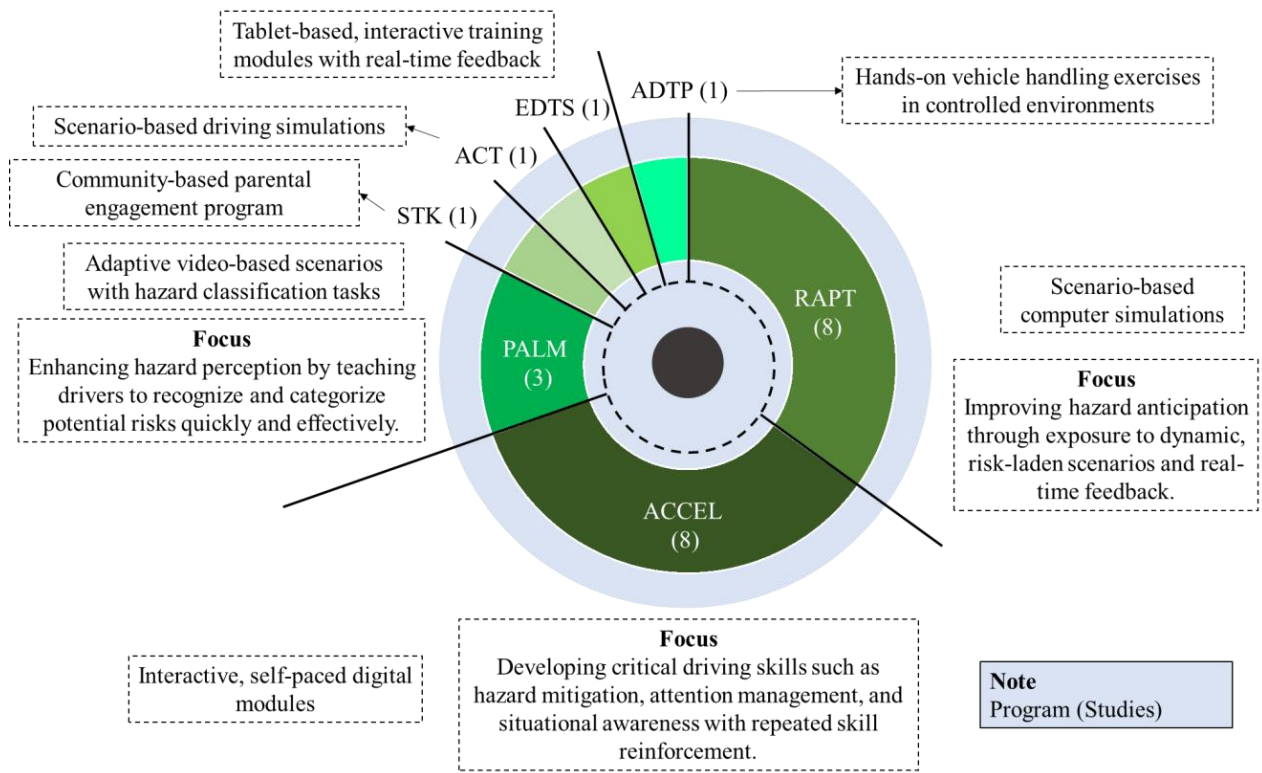

**Figure 5 Simulation and PC/Tablet-based programs**

*Risk Awareness Perception Training (RAPT)*

The RAPT program focuses on improving hazard perception by exposing novice drivers to high-risk driving scenarios in a computer-based training format. Participants interact with scenarios, receiving real-time feedback on their ability to scan for and anticipate latent hazards. Studies demonstrate that RAPT improves scanning behavior significantly, with hazard recognition increasing from 30% in untrained drivers to over 62% among those who completed the training. These effects were retained in follow-up evaluations up to six months later (Ahmadi et al., 2018; Pradhan et al., 2005). The training uses top-down and perspective views of roadway scenarios to highlight areas where risks may emerge, such as hidden pedestrians at intersections or vehicles merging unexpectedly (Yahoodik and Yamani, 2020).

*Virtual Reality Risk Awareness Perception Training (V-RAPT)*

V-RAPT is an extension of RAPT, delivered through virtual reality platforms like Oculus Rift. This immersive program presents drivers with a 360-degree environment to simulate real-world hazards more vividly. Evaluation results showed a significant improvement in hazard anticipation rates, with V-RAPT participants identifying 86.25% of latent hazards compared to 62.36% in the RAPT group (Agrawal et al., 2017). The enhanced realism in V-RAPT likely contributed to greater skill transfer from the virtual to real-world driving environments.

*Engaged Driver Training System (EDTS)*

The EDTS program is a tablet-based hazard anticipation and distraction management training tool. It builds on the success of RAPT but incorporates tactile and user-driven elements designed for mobile devices. EDTS has demonstrated significant short-term and long-term improvements in hazard awareness during both simulator and on-road tests. Teens who completed EDTS were more likely to identify and respond to latent hazards than those who underwent traditional or placebo training (Ahmadi et al., 2018; Zafian et al., 2016). The training includes interactive exercises with an error-based learning approach known as the "3M Method" (Mistake, Mitigation, and Mastery), which involves real-time feedback for improved learning retention (Ahmadi et al., 2018).

*Accelerated Curriculum to Create Effective Learning (ACCEL)*

ACCEL aims to accelerate the development of critical driving skills through a flexible, online training format. Participants can access the training on various platforms, including PCs, tablets, and smartphones, making it widely available for learners anywhere and anytime. Initial simulator-based evaluations demonstrated immediate improvements in crash-avoidance behaviors, with the effects enduring over time. Participants who received a "booster" training session performed even better in hazard anticipation during a second evaluation (Fisher et al., 2017; Glassman et al., 2024). ACCEL is particularly effective for female drivers, who showed greater improvements in risk awareness following training.

*Advanced Driver Training Programs (ADTPs)*

ADTPs combine classroom instruction with hands-on defensive driving exercises conducted on controlled tracks. These programs focus on vehicle handling, skid control, emergency stopping, and evasive maneuvers. Despite their popularity, results from evaluations have been mixed. For example, an ADTP evaluated in North Carolina did not demonstrate a statistically significant reduction in crash rates compared to traditional programs. Critics attribute these findings to overconfidence among participants, who may overestimate their skills after limited in-car experience (Wang et al., 2020).

*Driving Simulator-Based Training*

Simulator-based programs are widely used to replicate real-world driving conditions in a controlled setting. These simulators allow teens to practice risk-laden scenarios, such as navigating sharp curves, avoiding distracted driving, and maintaining proper lane discipline. Research shows that simulator training can reduce errors like road edge incursions and improve time-to-collision reactions (Campbell et al., 2016). However, studies indicate that simulator improvements do not always translate into long-term reductions in real-world crash risks unless supplemented with continuous training or real-life practice (Allen et al., 2012; Campbell et al., 2016; Pollatsek et al., 2011).

*Attention Training*

A recent longitudinal study examined the impact of a second exposure to hazard anticipation training. Participants who underwent a second round of ACCEL training exhibited improved hazard awareness and scanning accuracy during follow-up evaluations (Glassman et al., 2024). This approach underscores the importance of reinforcing learned skills to maintain and enhance long-term effectiveness, a key consideration for policymakers when designing driver education curricula. Across these technology-enhanced training methods, key features include interactive and immersive environments, real-time feedback, and scenario-based learning. Programs like V-RAPT and EDTS have demonstrated notable improvements in both short-term and long-term hazard perception, although the challenge of ensuring sustained behavior change in real-world driving remains. Advanced driver training programs with physical components are popular but require further research to determine their impact on crash rates beyond confidence-building exercises. Overall, technology-based methods offer scalable, accessible, and adaptable solutions to address the high-risk behaviors exhibited by novice teen drivers.

*Integration of Digital Applications in Teen Driver Education*

In addition to several in-person, virtual, and advanced training programs, there are various applications designed to support teen driving education and safety. RoadReady (roadreadyapp.com) helps teens track and log supervised driving hours, providing a comprehensive record for meeting state Graduated Driver Licensing (GDL) requirements. Similarly, Driving Tests (driving-tests.org) offers a range of online practice tests and learning modules to help teens prepare for their learner's permit exams through interactive quizzes and state-specific resources. Another app, Zutobi (zutobi.com), provides a gamified learning experience, combining bite-sized lessons with quizzes to improve knowledge retention for teen drivers across different licensing stages. For parental involvement, Safe2Drive (safe2drive.com) offers both driver education and defensive driving courses, often approved by state departments of motor vehicles. Additionally, Life360 (life360.com) focuses on safety by enabling location sharing, driving reports, and crash detection, ensuring real-time monitoring and alerts for teen drivers and their families. These apps are typically accessible via smartphones and web platforms, making them highly flexible and scalable solutions. Although evaluations indicate varying levels of success, many studies support their effectiveness in improving knowledge, tracking driving practice, and reinforcing safe driving behaviors when integrated with structured training programs.

**Table 4. Summary of Technology-Enhanced Training Methods for Teen Drivers.**

| Study | Key Findings/Summary |
|---|---|
| (Ahmadi et al., 2018) | Evaluated a tablet-based training program (EDTS) and found that it improved hazard perception skills in teen drivers both in the short term and six months after training. |
| (Glassman et al., 2024) | Investigated the impact of a "booster dose" of ACCEL training on hazard anticipation. Found that a second exposure significantly improved latent hazard anticipation in young drivers up to six months post-training. |
| (Pradhan et al., 2005) | Studied PC-based risk awareness training and found that trained novice drivers fixated more on risk-relevant areas in a driving simulator, indicating improved hazard perception. |
| (Fisher et al., 2016) | Examined the use of driving simulators as training and evaluation tools for novice drivers, highlighting their role in improving hazard anticipation and reducing risky behaviors. |
| (Allen et al., 2012) | Conducted a longitudinal study on simulator-based training for novice drivers and found that repeated exposure to hazardous scenarios improved long-term driving performance. |
| (Campbell et al., 2016) | Analyzed the efficacy of driving simulator training for teen drivers. Results showed no significant reduction in self-reported crashes or infractions, raising concerns about real-world applicability. |
| (Zafian et al., 2016) | Assessed the on-road effectiveness of a tablet-based training intervention (EDTS). Trained teens showed better hazard anticipation, but the effect was not statistically significant when parents were included in training. |

| (Fisher et al., 2017) | Developed the ACCEL program, a self-administered digital training tool for novice drivers. Found immediate post-training improvements in hazard anticipation, with effects persisting up to six months. |
|---|---|
| (Agrawal et al., 2017) | Evaluated a V-RAPT and found that VR-trained drivers had significantly better hazard anticipation than those using traditional RAPT. |

**Psychological and Cognitive Factors in Teen Driver Education**

Psychological and cognitive factors such as attention maintenance, hazard perception, and multitasking capabilities are critical determinants of teen driving performance and safety outcomes. Teen drivers, often described as "clueless" rather than "careless," struggle primarily due to cognitive limitations rather than deliberate risk-taking (Unverricht et al., 2019). Many educational programs have been developed to tackle these cognitive deficits, focusing on perceptual learning, visual scanning, and strategic decision-making.

The Forward Concentration and Attention Learning (FOCAL) program, for instance, focuses on enhancing attention maintenance skills by training drivers to limit off-road glances to less than two seconds while performing secondary tasks. In simulated driving scenarios, FOCAL-trained participants reduced their long glances by 24 percentage points compared to placebo-trained drivers and improved performance on both driving and in-vehicle tasks by 16 percentage points (Unverricht et al., 2019). This evidence supports the idea that FOCAL enhances multitasking abilities rather than merely reducing secondary task engagement. The RAPT program also addresses cognitive limitations by teaching drivers to detect latent hazards through repeated exposure to scenario-based simulations. Research has demonstrated sustained improvements in hazard anticipation, with RAPT-trained drivers increasing their correct hazard recognition rates by over 30% and retaining these gains for up to six months (Ahmadi et al., 2018; Pradhan et al., 2005). Similarly, V-RAPT program further enhances cognitive skill development through immersive 360-degree simulations, resulting in an 86.25% hazard anticipation rate (Agrawal et al., 2017).

Programs such as EDTS and ACCEL emphasize situational awareness and risk recognition through interactive learning modules. EDTS incorporates real-time error feedback and repetition to reinforce hazard awareness, showing positive long-term effects on attention and visual scanning behavior (Ahmadi et al., 2018; Zafian et al., 2016). ACCEL participants demonstrated sustained cognitive improvements in hazard anticipation, with a second training session significantly enhancing their performance compared to a single exposure (Fisher et al., 2017; Glassman et al., 2024). Driving simulators are also crucial for developing cognitive and perceptual skills in novice drivers. They provide opportunities to practice scenarios such as curve navigation and lane discipline in controlled environments. Although simulators improve performance metrics such as time-to-collision reactions, researchers emphasize the need for further evaluations to understand the long-term real-world impact of these cognitive training tools (Allen et al., 2012; Campbell et al., 2016).

These training programs collectively aim to strengthen novice drivers' perceptual-cognitive abilities by targeting key psychological factors contributing to crashes, such as visual attention lapses and inadequate hazard recognition. The integration of scenario-based exercises, feedback loops, and immersive environments has proven effective in improving both immediate and sustained driver behavior (Unverricht et al., 2019; Zafian et al., 2016). However, continuous reinforcement and follow-up training remain essential to ensure that these cognitive gains translate into safer driving practices over time(Glassman et al., 2024). Table 5 presents the summary of psychological and cognitive factors in teen driver education.

**Table 5. Psychological and Cognitive Factors in Teen Driver Education.**

| Study | Key Findings/Summary |
|---|---|
| (Fisher et al., 2017) | Evaluated the ACCEL training program, showing that it significantly improved hazard anticipation and attention maintenance. A second dose of training enhanced retention and effectiveness, particularly for female drivers. |
| (Agrawal et al., 2017) | Investigated V-RAPT for hazard anticipation. Found that VR training was significantly more effective in helping young drivers detect latent hazards compared to traditional methods. |
| (Ahmadi et al., 2018) | Examined the impact of a tablet-based training program on hazard perception. Results indicated that trained drivers demonstrated improved situation awareness and retained hazard anticipation skills for up to six months. |
| (Unverricht et al., 2019) | Assessed the FOCAL training program, which enhanced young drivers' ability to maintain attention on the roadway. Found that the program also improved multitasking skills, allowing drivers to manage cognitive load more effectively. |
| (Glassman et al., 2024) | Conducted a longitudinal study on ACCEL training and found that a second dose of training significantly improved hazard anticipation compared to a single exposure, supporting the need for reinforcement in cognitive training. |
| (Pradhan et al., 2005) | Evaluated the effectiveness of a computer-based RAPT. Found that trained novice drivers exhibited improved eye movement behavior and better fixation on risk-relevant areas in a driving simulator. |
| (Allen et al., 2012) | Investigated cognitive retention in novice drivers using different simulator configurations. Found that higher fidelity simulators led to better risk anticipation and improved cognitive retention over time. |
| (Campbell et al., 2016) | Studied the impact of simulator training on crash rates and driving infractions. Results showed no significant reduction in self-reported crashes, indicating a need for improved engagement and reinforcement in cognitive training. |
| (Zafian et al., 2016) | Evaluated a tablet-based training intervention for teen drivers. Found that trained participants exhibited improved hazard anticipation and reduced engagement in distracting behaviors while driving. |

**Parental and Peer Influence on Teen Driving Education**
Parental involvement is a key factor in shaping the driving behavior of teens, especially during the early stages of learning. Programs designed to enhance parental engagement have demonstrated significant improvements in safety outcomes by promoting communication, supervision, and reinforcement of safe driving behaviors. One prominent initiative is the Share the Keys (STK) program, implemented in New Jersey. STK is a community-based program that educates parents on the importance of their role in the GDL process. During a 90-minute orientation, parents and teens work together to develop contracts that address GDL compliance, curfews, and passenger restrictions. Over 85% of participating parents reported an increase in their ability to enforce these restrictions at home (Knezek et al., 2018). The Checkpoints program, supported by research from the University of Michigan Transportation Research Institute, employs a structured agreement between parents and teens to limit high-risk driving conditions. Parents who participated in Checkpoints imposed stricter limits on nighttime driving and peer passengers. Teens exposed to the program reported fewer incidents of risky driving, with 74% maintaining the agreement terms six months after licensure (Knezek et al., 2018). Checkpoints effectively combine active parental involvement with targeted education, ensuring sustained reductions in risky behavior.

The Parents Are the Key campaign, developed by the CDC, emphasizes the importance of ongoing parental involvement and provides tools for parents to communicate driving risks. However, unlike Checkpoints, the campaign's passive approach has faced criticism for its limited engagement, relying more on informational resources such as fact sheets and posters (Knezek et al., 2018). Studies have found that passive dissemination of educational materials has minimal impact unless paired with direct, structured engagement programs (Alderman et al., 2018). Parenting styles also significantly affect driving behavior. According to the National Young Driver Survey, authoritative parents who combine high support with clear rules are most successful in reducing risky behaviors, such as speeding, substance use, and distracted driving. Teens with authoritative parents were twice as likely to wear seat belts and half as likely to drive while angry compared to teens with uninvolved parents (Alderman et al., 2018). This emphasizes the importance of active parental supervision and role modeling in minimizing crash risks.

Programs like FOCAL have also highlighted the role of parents by encouraging structured supervised practice sessions. FOCAL-trained teens, guided by parents, demonstrated improved attention management and fewer distraction-related driving errors (Unverricht et al., 2019). Additionally, structured GDL programs, which often require parental involvement in supervised driving logs and permit application processes, have contributed to reductions in crash rates by as much as 38% in some regions. Peer influence, however, presents both risks and opportunities. Studies show that the presence of multiple teen passengers increases the likelihood of risky behaviors, particularly in male drivers. Programs that integrate peer-led discussions, such as those focusing on shared accountability for safety, have shown promise in countering negative peer influences. By promoting a culture of safe driving within social groups, these programs aim to mitigate risk factors associated with peer pressure (Alderman et al., 2018). The summary of parental and peer influence on teen driving education is presented in Table 6.

**Table 6. Parental and Peer Influence on Teen Driving Education.**

| Study | Key Findings/Summary |
|---|---|
| Knezek et al. (2018) | Evaluated the STK program, a behavioral-based parental training intervention. Found that parental engagement in the GDL process significantly influenced teen driver compliance with restrictions, with authoritative parenting styles being the most effective in reducing teen crash risks. |
| Unverricht et al. (2019) | Examined the role of peer influence in attention maintenance training (FOCAL). Found that teen drivers trained in attention maintenance were more resistant to peer distractions and maintained better focus on the road, suggesting cognitive training mitigates negative peer influence. |
| Alderman & Johnston (2018) | Investigated the impact of parental monitoring on teen driving behaviors. Found that authoritative parenting styles were associated with fewer risky driving behaviors, while uninvolved parenting correlated with increased crash risks and lower seat belt use. |

**Risk Awareness and Hazard Perception Training**

Risk awareness and hazard perception training are essential to address cognitive limitations in novice teen drivers. These programs target skills like anticipatory scanning, visual attention, and reaction timing to reduce crash risks, particularly in high-risk scenarios such as intersections, hidden hazards, and complex traffic conditions. The RAPT program has been widely studied and implemented to improve hazard recognition through scenario-based learning. RAPT trains drivers to scan the forward roadway and identify latent risks, such as obscured pedestrians and vehicles. Studies show that RAPT can increase hazard anticipation rates by over 30%, with participants demonstrating improved visual scanning behaviors that persist up to six months post-training (McDonald et al., 2015b; Unverricht et al., 2018). However, evaluations reveal that retention varies, with some decline in hazard detection after extended periods without reinforcement (McDonald et al., 2015a).

The Perceptual Adaptive Learning Module (PALM), developed by the AAA Foundation for Traffic Safety, uses adaptive, perceptual learning techniques to rapidly enhance novice drivers' hazard anticipation abilities. PALM presents video-based scenarios requiring participants to identify potential hazards and classify their causes. The program's effectiveness is demonstrated through significant improvements in time-to-glance and glance frequency metrics during simulator tests, particularly for complex events like mid-block crosswalks and lane transitions (Reyes and O'Neal, 2020). The ACCEL program complements PALM by offering strategic and tactical training on hazard anticipation, mitigation, and attention maintenance. In a recent simulator-based evaluation, ACCEL participants performed better in identifying and responding to dynamic hazards compared to control groups, particularly in scenarios involving hidden or sudden hazards. Simulator evaluations have highlighted the role of targeted training in improving both cognitive and performance-based measures. For example, PALM-trained drivers exhibited a 1.10 increase in hazard-related glances during post-training scenarios, indicating improved scanning behavior. Additionally, ACCEL-trained drivers displayed a significant 1.35-second increase in total glance time toward potential hazards, suggesting enhanced situational awareness. These programs also show that attentional control is critical for effective hazard perception. Studies found that trained drivers were more likely to allocate glances to critical areas, such as crosswalks and intersections, even when competing distractions were present. This ability to prioritize hazards over non-critical stimuli significantly reduces the risk of crashes in real-world driving conditions (Reyes and O'Neal, 2020). Overall, risk awareness and hazard perception training programs have demonstrated effectiveness in enhancing cognitive readiness and driving performance (Plumert et al., 2021; Reyes and O'Neal, 2020; Unverricht et al., 2018). Table 7 presents the summary of risk awareness and hazard perception training programs.

**Table 7. Summary of Risk Awareness and Hazard Perception Training Programs.**

| Study | Key Findings/Summary |
|---|---|
| (McDonald et al., 2015a) | Reviewed multiple hazard anticipation training programs, finding that while they improved immediate performance, long-term effects on crash reduction remain uncertain. |
| (Thomas et al., 2016) | Evaluated the RAPT program and found that trained males had a 23.7% lower crash rate, but there was no significant reduction for females. Training did not impact traffic violations. |
| (Unverricht et al., 2018) | Meta-analysis of 19 hazard anticipation training studies confirmed effectiveness in improving young drivers' latent hazard anticipation but emphasized the need for more on-road evaluations. |

| (Reyes and O'Neal, 2020) | Simulator-based evaluation of two training programs showed improved hazard recognition among novice drivers, with stronger effects for programs using interactive scenarios. |
|---|---|
| (Yahoodik and Yamani, 2020) | Found that RAPT-trained drivers maintained hazard anticipation performance even in complex environments with distractions, suggesting effective top-down attentional control. |
| (Plumert et al., 2021) | Extended evaluation of hazard anticipation training confirmed retention of skills over six months but noted variability in individual performance improvements. |
| (McDonald et al., 2015b) | Examined RAPT-3's impact on left-turn intersection behavior; found improvements in scanning for latent hazards but no significant change in actual crash rates. |
| Thomas et al. (2017) | Updated RAPT training with high-definition videos and interactive features, leading to significant improvements in hazard detection, sustained in post-training evaluations. |

**Long-Term Impact & Policy Implications of Driving Education Advancements**

Advancements in teen driver education have demonstrated promising safety improvements, particularly when integrated with GDL systems. Programs like Share the Keys (STK) and Checkpoints show sustained benefits through parental engagement, with over 85% of participants maintaining adherence to GDL restrictions and continued use of parent-teen contracts reducing risky driving incidents for up to six months (Knezek et al., 2018; Shell et al., 2015). Similarly, RAPT, PALM, and ACCEL have improved hazard perception in the short term but require periodic reinforcement to sustain long-term effectiveness (Reyes and O'Neal, 2020). Policy recommendations emphasize the need for ongoing training and evaluation to maintain critical cognitive skills and ensure lasting safety improvements. Additionally, policymakers advocate for increased access to simulation-based and adaptive training tools as part of standardized driver education programs.

**STRENGTHS, LIMITATIONS, AND LESSONS LEARNED**

The systematic review demonstrated significant strengths in its comprehensive coverage of various aspects of teen driver education. By including 29 studies spanning multiple methodologies, the research effectively captured a broad range of training interventions, from traditional programs to technology-enhanced methods. The inclusion of experimental studies, policy reviews, and longitudinal evaluations provided valuable insights into both short-term and long-term outcomes related to teen driver safety. Additionally, the classification of studies into thematic categories highlighted the diverse factors influencing novice drivers, including parental involvement, hazard perception, and cognitive skills.

However, several limitations were noted. The exclusion of 93 studies during the eligibility assessment due to focus on non-U.S. contexts, adult novice drivers, or unrelated content limited the scope of international comparisons. Furthermore, 55 reports could not be retrieved, which may have affected the comprehensiveness of the findings. Another challenge was the variability in study designs and reporting standards, making it difficult to directly compare outcomes across programs. In particular, a lack of long-term follow-up data in many studies limited the ability to assess sustained behavioral changes and crash risk reductions.

Despite these challenges, important lessons emerged from the review. The findings underscored the need for multi-faceted approaches that integrate traditional education with modern interventions such as simulation-based training and parental engagement programs. Programs like RAPT and STK demonstrated promising short-term improvements, though their long-term effectiveness depends on reinforcement and follow-up training. Additionally, addressing socioeconomic and geographic disparities in access to driver education was identified as a crucial policy consideration. These lessons provide a foundation for future research and policy development aimed at enhancing the effectiveness and equity of teen driver education programs.

**CONCLUSION**

The findings of this systematic review revealed key advancements and limitations in teen driver education programs between 2000 and 2024. Traditional education programs, including high school-based curricula and GDL initiatives, demonstrated effectiveness in imparting foundational driving skills but showed limited impact on long-term crash risk reduction. Emerging technology-enhanced interventions, such as RAPT, V-RAPT, and simulator-based training, provided measurable improvements in hazard anticipation and cognitive skills, though their effectiveness in real-world driving scenarios remains a topic for further investigation. Programs involving parental engagement, like Share the Keys (STK) and Checkpoints, highlighted the importance of structured supervision and communication in promoting compliance with GDL restrictions. However, barriers related to socioeconomic disparities, geographic accessibility, and policy inconsistencies continue to hinder equitable access to driver education. Collectively, these findings contribute to the growing body of literature by emphasizing the need for integrated approaches that blend traditional methods with innovative training tools and involve both parents and peers to enhance safety outcomes.

Future studies should explore the long-term impact of emerging driver education interventions, particularly those utilizing virtual reality, tablet-based training, and simulation technology. Longitudinal studies are needed to assess the sustained effects of these programs on crash reduction, risky driving behaviors, and skill retention over time. Additionally, there is a need for research on scaling and standardizing advanced training tools to promote broader access across diverse populations. Researchers should also investigate how these programs can better integrate into GDL policies and licensing systems to ensure consistency in training outcomes. Further exploration of the role of parental and peer influences on teen drivers' behavior, particularly in varying socioeconomic contexts, would provide valuable insights into enhancing program effectiveness. Lastly, mixed-methods studies incorporating both qualitative and quantitative approaches could provide a more comprehensive understanding of how novice drivers perceive and respond to training interventions.

From a practical perspective, policymakers and educators should consider enhancing traditional driver education programs by incorporating proven technology-based interventions such as RAPT, EDTS, and V-RAPT. Training curricula should emphasize hazard awareness and cognitive skills while ensuring continued engagement through booster sessions and follow-up training. Additionally, addressing disparities in access through targeted policy initiatives, such as subsidized training for low-income families and expanding training facilities in rural areas, is crucial for promoting equity in driver education. Programs that actively involve parents in the training process, such as STK, should be expanded and promoted to improve teen compliance with safety guidelines. Finally, continuous evaluation and refinement of these programs based on empirical evidence will be essential to ensure lasting improvements in teen driver safety and crash prevention.

**FUNDING**

This systematic review was conducted without any external funding or financial support.

**DECLARATION OF COMPETING INTEREST**

The authors declare that they have no known competing financial interests or personal relationships that could have appeared to influence the work reported in this paper.

**REFERENCES**

AAA, 2018. Orientation Sessions for Parents of Young Novice Drivers: An Assessment of U.S. Programs and Recommendations.

Agrawal, R., Knodler, M., Fisher, D.L., Samuel, S., 2017. Advanced virtual reality based training to improve young drivers' latent hazard anticipation ability, in: Proceedings of the Human Factors and Ergonomics Society Annual Meeting. SAGE Publications Sage CA: Los Angeles, CA, pp. 1995–1999.

Ahmadi, N., Katrahmani, A., Romoser, M.R., 2018. Short and long-term transfer of training in a tablet-based teen driver hazard perception training program, in: Proceedings of the Human Factors and Ergonomics Society Annual Meeting. SAGE Publications Sage CA: Los Angeles, CA, pp. 1965–1969.

Alderman, E.M., Johnston, B.D., Breuner, C., Grubb, L.K., Powers, M., Upadhya, K., Wallace, S., Hoffman, B.D., Quinlan, K., Agran, P., others, 2018. The teen driver. Pediatrics 142.

Allen, R.W., Park, G.D., Cook, M.L., Fiorentino, D., 2012. Simulator training of novice drivers: a longitudinal study. Advances in Transportation Studies.

Campbell, B.T., Borrup, K., Derbyshire, M., Rogers, S., Lapidus, G., 2016. Efficacy of driving simulator training for novice teen drivers. Conn Med 80, 291–296.

Curry, A.E., García-España, J.F., Winston, F.K., Ginsburg, K., Durbin, D.R., 2012. Variation in teen driver education by state requirements and sociodemographics. Pediatrics 129, 453–457.

Curry, A.E., Peek-Asa, C., Hamann, C.J., Mirman, J.H., 2015. Effectiveness of parent-focused interventions to increase teen driver safety: A critical review. Journal of Adolescent Health 57, S6–S14.

Das, S., Minjares-Kyle, L., Wu, L., Henk, R.H., 2019. Understanding crash potential associated with teen driving: survey analysis using multivariate graphical method. Journal of safety research 70, 213–222.

Edwards, M., 2001. Standards for novice driver education and licensing. E-CIRCULAR 22.

Fisher, D.L., Dorn, L., 2016. The training and education of novice teen drivers, in: Handbook of Teen and Novice Drivers. CRC Press, pp. 289–310.

Fisher, D.L., Young, J., Zhang, L., Knodler, M., Samuel, S., 2017. Accelerating teen driver learning: Anywhere, anytime training. Behavior Research Methods 37, 379–384.

Foss, R.D., Masten, S.V., Goodwin, A.H., O'Brien, N.P., TransAnalytics, L., others, 2012. The role of supervised driving requirements in graduated driver licensing programs. United States. National Highway Traffic Safety Administration.

GAO, 2010. Teen Driver Safety: Additional Research Could Help States Strengthen Graduated Driver Licensing Systems (No. GAO-10-544).

Gesser-Edelsburg, A., Guttman, N., 2013. "Virtual" versus "actual" parental accompaniment of teen drivers: A qualitative study of teens' views of in-vehicle driver monitoring technologies. Transportation research part F: traffic psychology and behaviour 17, 114–124.

Glassman, J., Yahoodik, S., Samuel, S., Young, J., Knodler, M.K., Zhang, T., Zafian, T., Fisher, D.L., Yamani, Y., 2024. Booster dose of attention training program for young novice drivers: a longitudinal driving simulator evaluation study. Human factors 66, 933–953.

Goode, N., Salmon, P.M., Lenné, M.G., 2013. Simulation-based driver and vehicle crew training: applications, efficacy and future directions. Applied ergonomics 44, 435–444.

Hossain, M.M., Zhou, H., Das, S., Sun, X., Hossain, A., 2023. Young drivers and cellphone distraction: Pattern recognition from fatal crashes. Journal of Transportation Safety & Security 15, 239–264.

Hossain, M.M., Zhou, H., Rahman, M.A., Das, S., Sun, X., 2022. Cellphone-distracted crashes of novice teen drivers: Understanding associations of contributing factors for crash severity levels and cellphone usage types. Traffic injury prevention 23, 390–397.

IIHS, 2023. Fatality Facts 2022 Yearly Snapshot.

Knezek, C.M., Polirstok, S., James, R., Poedubicky, G., 2018. Effects of Share the Keys (STK), a comprehensive, behavioral-based training program for parents and new teen drivers in New Jersey. Transportation research part F: traffic psychology and behaviour 56, 156–166.

Lee, S.E., Simons-Morton, B.G., Klauer, S.E., Ouimet, M.C., Dingus, T.A., 2011. Naturalistic assessment of novice teenage crash experience. Accident Analysis & Prevention 43, 1472–1479.

Lonero, L., Mayhew, D., 2010. Large-scale evaluation of driver education review of the literature on driver education evaluation 2010 update. AAA Foundation for Traffic Safety, Washington, DC.

Masten, S.V., Foss, R.D., 2010. Long-term effect of the North Carolina graduated driver licensing system on licensed driver crash incidence: A 5-year survival analysis. Accident Analysis & Prevention 42, 1647–1652.

Mayhew, D.R., Vanlaar, G., Robertson, R.D., others, 2024. Driver Education and Training Promising Practices: A Systemic Literature Review.

McCartt, A.T., Mayhew, D.R., Braitman, K.A., Ferguson, S.A., Simpson, H.M., 2009. Effects of age and experience on young driver crashes: review of recent literature. Traffic injury prevention 10, 209–219.

McDonald, C.C., Goodwin, A.H., Pradhan, A.K., Romoser, M.R., Williams, A.F., 2015a. A review of hazard anticipation training programs for young drivers. Journal of Adolescent Health 57, S15–S23.

McDonald, C.C., Kandadai, V., Loeb, H., Seacrist, T., Lee, Y.-C., Bonfiglio, D., Fisher, D.L., Winston, F.K., 2015b. Evaluation of a risk awareness perception training program on novice teen driver behavior at left-turn intersections. Transportation research record 2516, 15–21.

Moher, D., Liberati, A., Tetzlaff, J., Altman, D.G., 2009. Preferred Reporting Items for Systematic Reviews and Meta-Analyses: The PRISMA Statement. Journal of Clinical Epidemiology 62, 1006–1012. https://doi.org/10.1016/j.jclinepi.2009.06.005

Muttart, J.W., Agrawal, R., Ebadi, Y., Samuel, S., Fisher, D.L., 2017. Evaluation of a training intervention to improve novice drivers' hazard mitigation behavior on curves, in: Driving Assessment Conference. University of Iowa.

National Highway Traffic Safety Administration, 2022. Traffic Safety Facts Annual Report Tables. National Highway Traffic Safety Administration.

NHTSA, 2009. Novice teen driver education and training administrative standards. National Highway Traffic Safety Administration (NHTSA): Washington, DC, USA.

O'Neill, B., 2020. Driver education: how effective? International Journal of Injury Control and Safety Promotion 27, 61–68. https://doi.org/10.1080/17457300.2019.1694042

Plumert, J.M., Reyes, M., O'Neal, E.E., Vecera, S., Allen, S., McGehee, D.V., others, 2021. Extended evaluation of training programs to accelerate hazard anticipation skills in novice teens drivers. Safety Research Using Simulation (SAFER-SIM) University Transportation Center.

Pollatsek, A., Vlakveld, W., Kappé, B., Pradhan, A., Fisher, D.L., 2011. Driving simulators as training and evaluation tools: Novice drivers. D. Fisher, M. Rizzo, J. Caird, & J. Lee, Handbook of driving simulation for Engineering, Medicine and Psychology 4.

Pradhan, A.K., Fisher, D.L., Pollatsek, A., 2005. The effects of PC-based training on novice drivers' risk awareness in a driving simulator, in: Driving Assessment Conference. University of Iowa.

Reyes, M.L., O'Neal, E., 2020. A simulator-based evaluation of two hazard anticipation training programs for novice drivers.

Ryerson, M., Davidson, J., Wu, J.S., Feiglin, I., Winston, F., 2022. Identifying community-level disparities in access to driver education and training: Toward a definition of driver training deserts. Traffic injury prevention 23, S14–S19.

Ryerson, M.S., Dong, X., Wu, J.S., Walshe, E.A., Winston, F.K., others, 2024. Disparities in Access to Driver Education for Teens as a Health and Mobility Equity Issue.

Shell, D.F., Newman, I.M., Córdova-Cazar, A.L., Heese, J.M., 2015. Driver education and teen crashes and traffic violations in the first two years of driving in a graduated licensing system. Accident Analysis & Prevention 82, 45–52.

Thomas, F.D., Blomberg, R.D., Fisher, D.L., others, 2012. A fresh look at driver education in America (No. DOT HS 811 543). United States. National Highway Traffic Safety Administration.

Thomas, F.D., Korbelak, K.T., Divekar, G., Blomberg, R.D., Romoser, M.R., Fisher, D.L., others, 2017. Evaluation of an updated version of the risk awareness and perception training program for young drivers. United States. Department of Transportation. National Highway Traffic Safety ….

Thomas, F.D., Rilea, S., Blomberg, R.D., Peck, R.C., Korbelak, K.T., others, 2016. Evaluation of the safety benefits of the risk awareness and perception training program for novice teen drivers. Dunlap and Associates, Inc.

Unverricht, J., Samuel, S., Yamani, Y., 2018. Latent hazard anticipation in young drivers: Review and meta-analysis of training studies. Transportation research record 2672, 11–19.

Unverricht, J., Yamani, Y., Yahoodik, S., Chen, J., Horrey, W.J., 2019. Attention maintenance training: Are young drivers getting better or being more strategic?, in: Proceedings of the Human Factors and Ergonomics Society Annual Meeting. SAGE Publications Sage CA: Los Angeles, CA, pp. 1991–1995.

Wang, Y.C., Foss, R.D., O'Brien, N.P., Goodwin, A.H., Harrell, S., 2020. Effects of an advanced driver training program on young traffic offenders' subsequent crash experience. Safety Science 130, 104891.

Williams, A.F., 2017. Graduated driver licensing (GDL) in the United States in 2016: A literature review and commentary. Journal of safety research 63, 29–41.

Yahoodik, S., Yamani, Y., 2020. Attentional control in young drivers: Does training impact hazard anticipation in dynamic environments?, in: Proceedings of the Human Factors and Ergonomics Society Annual Meeting. SAGE Publications Sage CA: Los Angeles, CA, pp. 1986–1990.

Zafian, T.M., Samuel, S., Coppola, J., O'Neill, E.G., Romoser, M.R., Fisher, D.L., 2016. On-road effectiveness of a tablet-based teen driver training intervention, in: Proceedings of the Human Factors and Ergonomics Society Annual Meeting. SAGE Publications Sage CA: Los Angeles, CA, pp. 1926–1930.